\documentclass[acmsmall]{acmart}

\usepackage{longtable}
\usepackage{colortbl}
\usepackage{multirow}
\usepackage{subcaption}
\usepackage{csquotes}
\usepackage{color}
\usepackage{tabularray}

\acmArticle{CSCW258}

\received{January 2024}
\received[revised]{October 2024}
\received[accepted]{February 2025}

\setcopyright{cc}
\setcctype{by}
\acmJournal{PACMHCI}
\acmYear{2025} \acmVolume{9} \acmNumber{7} \acmArticle{258} \acmMonth{11} \acmPrice{}\acmDOI{10.1145/3757439}

\NewDocumentCommand{\pquote}{+O{} +m}{\blockquote[#1]{\textit{#2}}}

\newcommand{\elide}[1]{\textelp{}} % produces ... instead of argument

\newcommand{\one}[1]{\textcolor{black}{#1}}
\newcommand{\textcite}[1]{\citeauthor{#1}~\cite{#1}}

\begin{document}

\title{"When I Lost It, They Dragged Me Out": How \textit{Care Encounters} Empower Marginalized Young Adults' Mental Health Aspiration and Care-Seeking}

\author{Jiaying "Lizzy" Liu}
\affiliation{%
  \institution{School of Information, University of Texas at Austin}
  \city{Austin}
  \country{USA}}
\orcid{0000-0002-5398-1485}
\email{jiayingliu@utexas.edu}

\author{Yan Zhang}
\affiliation{%
  \institution{School of Information, University of Texas at Austin}
  \city{Austin}
  \country{USA}
}
\orcid{0000-0002-1130-0012}
\email{yanz@utexas.edu}
\renewcommand{\shortauthors}{Jiaying "Lizzy" Liu \& Yan Zhang}
\renewcommand{\shorttitle}{How Care Encounters Empower Marginalized Young Adults' Mental Health Aspiration and Care-Seeking}

\begin{abstract}
Mental health care-seeking among marginalized young adults has received limited attention in CSCW research. Through in-depth interviews and visual elicitation methods with 18 diverse U.S. participants, our study reveals how marginalized identities shape mental health care-seeking journeys, often characterized by \textit{low aspirations} and \textit{passive care-seeking} influenced by lived experiences of marginalization. However, we found the transformative function of  "\textit{care encounters}"—serendipitous interactions with mental health resources that occur when individuals are not actively seeking support. These encounters serve as critical turning points, catalyzing shifts in aspiration and enabling more proactive care-seeking behaviors. 
Our analysis identifies both the infrastructural conditions that enable transformative care encounters and the \textit{aspiration breakdowns} that impede care-seeking processes. 
This work makes conceptual contributions by supplementing traditional motivation-based care-seeking models with a reconceptualization of "care encounters" that accounts for the infrastructural and serendipitous nature of mental health access. We advance understanding of how marginalized identity uniquely influences care-seeking behaviors while providing actionable design implications for embedding technology-mediated "care encounters" into socio-technical interventions that can better support mental health care access for vulnerable populations.

%Their impact operates through three mechanisms: tangible assistance, supportive discourse, and social connection building.
%Our findings suggest strategies for intentionally engineering care encounters that can help overcome passive behavioral patterns and strengthen marginalized young adults' engagement with mental health resources.
\end{abstract}

\begin{CCSXML}
<ccs2012>
   <concept>
       <concept_id>10003120.10003121.10011748</concept_id>
       <concept_desc>Human-centered computing~Empirical studies in HCI</concept_desc>
       <concept_significance>500</concept_significance>
       </concept>
   <concept>
       <concept_id>10003120.10003121.10003126</concept_id>
       <concept_desc>Human-centered computing~HCI theory, concepts and models</concept_desc>
       <concept_significance>500</concept_significance>
       </concept>
   <concept>
       <concept_id>10003120.10003121.10003122.10003334</concept_id>
       <concept_desc>Human-centered computing~User studies</concept_desc>
       <concept_significance>500</concept_significance>
       </concept>
 </ccs2012>
\end{CCSXML}

\ccsdesc[500]{Human-centered computing~Empirical studies in HCI}
\ccsdesc[500]{Human-centered computing~HCI theory, concepts and models}
\ccsdesc[500]{Human-centered computing~User studies}

\keywords{Help-Seeking, Socio-technical ecosystem, Pathway to Care, Identity, Intersectionality, Health Information Behavior, Social Ecological Theory}

\maketitle

\section{Introduction}
The mental health crisis continues to escalate both in the United States and globally. According to the National Institute of Mental Health \cite{national_institute_of_mental_health_mental_2023}, approximately 57.8 million U.S. adults lived with a mental illness in 2021. Young adults aged 18-25 demonstrate the highest prevalence of mental health concerns (33.7\%), yet paradoxically exhibit the lowest rate of seeking outpatient mental health treatment. This disparity becomes even more pronounced among young adults with marginalized identities, particularly racial minorities and LGBTQ+ individuals, who demonstrate significantly lower care-seeking behaviors \cite{hunt_mental_2010}.

A substantial body of HCI and CSCW has addressed mental health support through various digital interventions, such as self-tracking tools \cite{wiljer_effects_2020, alqahtani_co-designing_2021, petelka_being_2020} and peer support platforms \cite{wong_postsecondary_2021, toscos_college_2018, zhang_online_2018}, targeting the general population. However, researchers have increasingly recognized that certain technological solutions may prove inadequate for marginalized populations, who often face limited access to technological resources \cite{hickey_smart_2021, miller_awedyssey_2023}. Moreover, these populations encounter multifaceted structural barriers in adopting such tools due to their unique cultural, social, and economic circumstances. For instance, \textcite{tachtler_unaccompanied_2021} identified specific micro and macro-level barriers that impede unaccompanied immigrant children's access to and utilization of mental health mobile applications.

Existing research has documented significantly low technology adoption rates among marginalized young adults seeking mental health support \cite{gulliver_perceived_2010}. To design ecologically valid socio-technical interventions, it is crucial to understand the lived experiences shaping their care-seeking journeys. While prior work has examined various barriers of marginalized communities to access mental health care \cite{shahid_asian_2021, soubutts_challenges_2024}, there remains a critical gap in understanding how perceived marginalized identity influences care-seeking practices. This research addresses this gap by investigating two research questions:

\begin{itemize}
    \item How do marginalized young adults navigate their care-seeking journey?
    \item How does the socio-technical ecosystem of mental health care resources facilitate or hinder care-seeking?
\end{itemize}

We interviewed 18 U.S. young adults with diverse marginalized identities, such as LGBTQ+, racial minorities, first-generation college students, and homeless. Our analysis reveals patterns of low aspiration in participants' care-seeking behaviors, demonstrating how their marginalized positions and identities constrain access to care and shape their perceptions of mental health resources. These limitations not only restrict their care expectations but also create isolation from support systems. In summary, this study makes the following contributions to CSCW, HCI, and ICT4D.

Importantly, we identified opportunities in fostering aspirations and empowering proactive care-seeking through "care encounters" - serendipitous interactions that provide marginalized young adults with tangible assistance, supportive discourse, and social connection. Our analysis examines how institutional and infrastructural designs facilitate these transformative encounters. However, we also identified two critical barriers within the existing socio-technical ecosystem that can impede transitions toward proactive care-seeking: challenges in establishing trust and limitations in empowering individual agency.

\begin{itemize}
\item We reveal through in-depth qualitative analysis how marginalized identity shapes and constrains young adults' mental health care-seeking journeys.
\item We identify and examine mechanisms of how "care encounters" transform marginalized young adults' aspirations and care-seeking behaviors, illuminating the infrastructural conditions that enable these pivotal interactions.
\item We propose design strategies for engineering effective care encounters considering the unique socio-technical ecosystem of marginalized young adults, addressing multiple levels of mental health support.
%: from individual technologies to community resources and institutional systems.
\end{itemize}

%\textbf{Terminology}. We acknowledge the limitations of the term of marginalized populations. We refer marginalized as a situation rather with the respect to their agency. 

\section{Related Work}
We first review the literature concerning general behavior models around mental health care-seeking and then center on the practices of marginalized populations. We then elaborate on the ecological view of mental health care-seeking which serves as an overarching methodology approach of this study. 

\subsection{Behavioral Modals around Mental Health Care-Seeking} 
Care-seeking is also referred to as help-seeking, denoting people's behavior of actively obtaining assistance from various sources \cite{rickwood_conceptual_2012}. The assistance includes suggestions, information, and treatment that serve pragmatic purposes and also social and emotional support and understanding \cite{lachmar_mydepressionlookslike_2017}. 

One of the earliest models focused on formal help-seeking in the medical system. \textcite{huxley_mental_1996} outlined a five-layer pathway that a patient may need to navigate through to get care: Potential patients entered the pathway from general practitioners, via whom they were referred to primary care, then secondary psychiatric care, and ended with admission to the hospital. However, with its focus on patients' trajectory within medical institutions, this model does not sufficiently consider contextual factors such as socio-economic status, cultural influences, or the role of informal support networks \cite{liu2025review}. Subsequent studies expanded on this model by identifying diverse entry points to formal mental health care that exist outside the medical system, including consultations with traditional or religious healers \cite{gater_pathways_1991}, interventions by law enforcement \cite{bhui_mental_2002}, and engagement with national crisis helplines \cite{pendse_can_2021}.

Moving beyond medical institutions, Rickwood's help-seeking framework \cite{rickwood_young_2005} sketches four key stages in the help-seeking process: (1) recognizing symptoms and realizing the need for assistance, (2) expressing symptoms and signaling the need for support, (3) identifying available and accessible sources of help, and (4) reaching out to the chosen source, contingent on the individual's willingness to disclose their difficulties. While comprehensive in scope, this framework presents these stages as static categories, potentially overlooking the dynamic and iterative nature of help-seeking over time.

Seeking help for mental health is not a quick and simple decision; rather, it is a complex and dynamic process dubbed with stigmatization and solitary feelings \cite{lannin_does_2016}. In the process, people constantly negotiate with themselves "if seeking help is necessary" and "when and how to seek help". \textcite{biddle_explaining_2007} delineated this struggle through the Cycle of Avoidance (COA) model, depicting the tensions involved in making sense of, accepting, and avoiding mental distress. They identified several key actions that postpone help-seeking, including normalizing symptoms, proposing alternative explanations, accommodating increasingly severe distress, pushing the threshold between “normal” and “real” distress, and delaying help-seeking. Other studies also pinpointed similar cognitive strategies such as utilizing self-resilience, denying the effects of professional help, and problematizing help-seeking \cite{martinez-hernaez_non-professional-help-seeking_2014, abavi_exploration_2020}. 

While these models capture the complexities of mental health care-seeking behaviors and stages, they do not examine how individuals interact with specific health resources and technologies—insights that could directly inform the design of future interventions. Furthermore, these models primarily draw from general population studies, leaving unclear how marginalized communities seek mental health care.

\subsection{Marginalized Populations' Care-Seeking Behaviors}
HCI, CSCW, and ICT4D scholars have increasingly focused their attention on marginality and the empowerment of society's most vulnerable populations \cite{chordia_social_2024, pearce_socially-oriented_2020, wyche_learning_2012, devito_social_2019, pendse_mental_2019}. Marginalization manifests in multiple dimensions, encompassing economic hardship, gender discrimination, ethnic exclusion, geographical isolation, educational disparities, and political disenfranchisement \cite{pal_marginality_2013}. These structural inequalities profoundly shape individuals' aspirations and life trajectories, with intersecting identities of race, class, gender, and disability \cite{das_studying_2023} creating complex barriers that diminish both the intention and capacity to seek care \cite{crenshaw_mapping_2013}.

The reality of limited resources presents immediate and tangible barriers for marginalized populations seeking healthcare services. Evidence demonstrates that individuals in remote rural areas face reduced access to essential services, such as helplines \cite{pendse_can_2021}, and utilize significantly fewer health services \cite{williams_analysis_2021}. This perception of resource scarcity may also contribute to a deeper pattern of symptom normalization and treatment avoidance \cite{shahid_asian_2021}. For example, \textcite{pendse_marginalization_2023} found that online mental health narratives from people in low-resource areas tend to emphasize somatic expressions of distress through body-focused language and have their experiences stigmatized and invalidated. Such evidence suggests that the impact of marginality extends beyond mere resource limitations, manifesting in more nuanced ways in people's identity perception and throughout the care-seeking process.

HCI research has documented the complex interactions between individuals with marginalized identities and mental health technologies \cite{oguamanam2023intersectional}, revealing significant cultural and social barriers to care. For instance, \textcite{bhattacharjee_whats_2023} illuminated how Indian users express profound skepticism toward talk therapy rooted in Western psychological paradigms, highlighting fundamental cultural disconnects in mental health interventions. Studies also demonstrated how perceived discrimination based on marginalized identities actively deters individuals from seeking clinical care \cite{lu_barriers_2021, gulliver_perceived_2010}. These negative experiences create lasting psychological barriers to healthcare access \cite{liu_exploring_2024}, initiating self-reinforcing cycles that further alienate individuals from mental health resources. The impact of these barriers extends into digital spaces, as \textcite{feuston_everyday_2019} documented how social media algorithms and platform affordances shape narratives around mental illness. These technological mechanisms, influenced by community norms that equate mental illness with deviance, inadvertently amplify and perpetuate stigma in online spaces, creating additional obstacles to care-seeking behaviors.

However, these studies mostly examined marginalized populations' interactions with specific technologies, and there remains a critical gap in understanding how individuals with marginalized identities navigate their broader care-seeking journey and engage with diverse available resources \cite{robards_how_2018}. This comprehensive understanding is crucial for developing culturally sensitive designs that align with ecological systems \cite{burgess_i_2019}, cultural values \cite{li_sunforum_2016}, and community preferences \cite{das_studying_2023}.

%Our study addresses this gap by investigating how marginalized populations negotiate the complex interplay between their personal identity, disadvantaged social position, and the healthcare system in their pursuit of care. This research aims to inform more inclusive and effective healthcare interventions that acknowledge and accommodate the unique challenges faced by marginalized communities.
%\cite{Seeking in Cycles: How Users Leverage Personal Information Ecosystems to Find Mental Health Information}
%People's real-world health resource utilization  \cite{andersen_behavioral_1968,andersen_revisiting_1995,babitsch_re-revisiting_2012} are enacted in interacting with complicated situations. 
\subsection{Ecological Approach for Technology-Mediated Care-Seeking}
Recent HCI scholarship has expanded beyond studying users' interactions with isolated personal information systems to examine them within complex social ecologies \cite{murnane_personal_2018, siddiqui_exploring_2023, ongwere_challenges_2022}. For example, \textcite{wong_mental_2023} examined how workplace environments shape and limit engineers' engagement with mental health resources.  \textcite{theofanopoulou_exploring_2022} studied tangible toys as mediators for parent- child emotional communication in family settings.
Bronfenbrenner's social systems framework \cite{bronfenbrenner1979ecology} has been applied in such HCI research to examine the layered context of health resource interactions. Researchers have used its components—microsystem, macrosystem, and chronosystem—to study social influences on technology-mediated mental health management \cite{burgess_i_2019, tachtler_unaccompanied_2021, murnane_personal_2018}.

This ecological perspective has particular relevance for marginalized populations who face distinct socio-economic considerations \cite{kaziunas_precarious_2019}. For instance, \textcite{tachtler_unaccompanied_2021} demonstrated how macro-level factors, particularly resettlement policies, constrain unaccompanied migrant youths' engagement with mental health mobile applications. Our study extends this ecological approach to examine how marginalized young adults navigate mental health resources within their lived contexts.

We draw on Social Ecological Theory \cite{stokols_translating_1996} to examine young adults' mental health navigation. The framework by \textcite{mcleroy_ecological_1988} identifies five levels affecting health behavior: individual characteristics, social relationships, organizational factors, community characteristics, and societal factors, including physical, social, and political environments \cite{mccloskey_principles_2011}. Following prior studies \cite{sallis_ecological_2008, trace_information_2023}, we employ the version adapted by the Centers for Disease Control and Prevention \cite{centers_for_disease_social-ecological_2022} to examine individuals' interactions with care resources and technologies within their ecosystem, as detailed in Section \ref{sec: analysismethod}. 

Through this ecological lens, we investigate how marginalized young adults navigate digital platforms, institutional services, and informal support networks within their socioeconomic constraints, aiming to identify barriers and enablers that can inform the design of mental health technologies and services \cite{c_feasibility_2022, wong_postsecondary_2021}.

\section{Methods}
We combined in-depth interviews with visual elicitation techniques to understand the lived experiences of marginalized young adults’ mental health care-seeking practices.
\subsection{Recruitment and Participants}
We disseminated recruitment messages on the listserv of a southern university in America as previous studies suggested that university students face high and multifaceted pressure \cite{hunt_mental_2010}. To recruit young adults from diverse backgrounds, we sourced participants through Reddit, aiming to include individuals from marginalized communities who are often underrepresented in university-based studies. Following Reddit's community guidelines, we posted recruitment messages in 12 regional marketplace and job-focused subreddits with large memberships across the United States, including r/SanAntonioJobs and r/NewOrleansMarketplace.

Interested participants ages 18-25 were invited to finish a screening questionnaire where we asked for their demographic information, such as age, gender, education, marginalized identities (e.g., first-generation college students, low-income, LGBTQ+, race minority), mental illness diagnosis, help-seeking practices (e.g., previously used resources and technologies such as family, friends, social media, and apps), and an open-ended question about the most recent/impressive mental health help-seeking experience. We also include the PHQ-4 \cite{kroenke_ultra-brief_2009}, a validated screening tool for depression and anxiety to get a rough estimate of their current mental health status. We intentionally chose participants to maximize the sample demographic diversity and mental health status.
Following the theoretical sampling strategy \cite{charmaz2000grounded}, our recruitment process proceeded in parallel with data analysis to include participants with diverse help-seeking practices, including varied preferred sources and both satisfying and dissatisfying experiences. For instance, we noticed the prominent role of resources actively reaching out to marginalized young adults based on the first twelve interviews. To collect more cases and further understand the care-outreaching, the team carefully selected subsequent participants who reported relevant experiences in the screening survey and invited them to join the interview.
%It consists of four questions asking about the frequency of feelings of "nervous, anxious or on edge", "not being able to stop or control worrying," "little interest or pleasure in doing things," and "feeling down, depressed, or hopeless" over the last two weeks. 

As shown in Table \ref{tab:participants}, 18 participants were interviewed (7 females, 9 males, and 2 trans/non-binary; 6 White, 5 Black, 4 Asian, and 3 Hispanic). Half of the participants were still in university at the time of the interviews, 6 had graduated from colleges, and 3 had no college education. Thirteen participants had formal diagnoses of mental illness, mainly depression and anxiety. Their anxiety and depression status indicated by PHQ-4 ranged from mild (3-5), moderate (6-8) to severe (9-12). 

\begin{table}[h]
\caption{Participants Information}
\label{tab:participants}
\resizebox{\columnwidth}{!}{%
\begin{tabular}{|l|l|l|l|l|l|l|l|l|}
\hline
\textbf{ID} &
  \textbf{Age} &
  \textbf{Sex} &
  \textbf{Education} &
  \textbf{Race} &
  \textbf{Diagnosis} &
  \textbf{PHQ4} &
  \textbf{Resources Used in Mental Health Help-seeking} \\ \hline
P01 &
  20 &
  F &
  Some college &
  Black &
  depression &
  \cellcolor[HTML]{8AC97D}5 &
  professional help, search engine, friend, family \\ \hline
P02 &
  24 &
  M &
  High school &
  Black &
  depression &
  \cellcolor[HTML]{FFEB84}8 &
  social media, online   communities, friends, professional help \\ \hline
P03 &
  20 &
  F &
  Some college &
  Asian &
  no &
  \cellcolor[HTML]{63BE7B}4 &
  social media, search engine, friends, family, mobile apps \\ \hline
P04 &
  23 &
  M &
  Some college &
  White &
  Anxiety &
  \cellcolor[HTML]{F8696B}12 &
  \begin{tabular}[c]{@{}l@{}}Family, friends, professionals, online communities, services in the local \\ community\end{tabular} \\ \hline
P05 &
  22 &
  M &
  8 through 11 years &
  Black &
  anxiety &
  \cellcolor[HTML]{F8696B}12 &
  family, friend, professional, social media, online communities, telehealth \\ \hline
P06 &
  22 &
  F &
  Some college &
  Asian &
  depression &
  \cellcolor[HTML]{FFEB84}8 &
  telehealth, professional help \\ \hline
P07 &
  21 &
  F &
  Some college &
  Hispanic &
  \begin{tabular}[c]{@{}l@{}}Depression, \\ Anxiety\end{tabular} &
  \cellcolor[HTML]{63BE7B}4 &
  Friends, professors, people with   similar experiences \\ \hline
P08 &
  23 &
  F &
  College graduate &
  Asian &
  Depression &
  \cellcolor[HTML]{D8DF81}7 &
  \begin{tabular}[c]{@{}l@{}}Family, friends, people with similar experiences, Social media, \\ online mental health forums or communities, Search engines, \\ Teletherapy services\end{tabular} \\ \hline
P09 &
  24 &
  M &
  College graduate &
  White &
   &
  \cellcolor[HTML]{8AC97D}5 &
  Family, friends, Reddit \\ \hline
P10 &
  24 &
  M &
  College graduate &
  Asian &
  no &
  \cellcolor[HTML]{63BE7B}4 &
  Hotline, family \\ \hline
P11 &
  22 &
  M &
  Some college &
  White &
  Depression &
  \cellcolor[HTML]{FCAA78}10 &
  Family, professionals, social media, search engine, teletherapy \\ \hline
P12 &
  20 &
  F &
  Some college &
  White &
  Depression &
  \cellcolor[HTML]{F8696B}12 &
  \begin{tabular}[c]{@{}l@{}}Family, professionals, people with similar experiences, social media, \\ online communities, teletherapy, hotlines, search engines\end{tabular} \\ \hline
P13 &
  22 &
  M &
  Some college &
  White &
  Depression &
  \cellcolor[HTML]{FECB7E}9 &
  Family, friends, search engines, mobile applications \\ \hline
P14 &
  21 &
  F &
  College graduate &
  Hispanic &
  Depression &
  \cellcolor[HTML]{8AC97D}5 &
  Professionals, online communities \\ \hline
P15 &
  25 &
  \begin{tabular}[c]{@{}l@{}}Trans\end{tabular} &
  Some college &
  Hispanic &
  \begin{tabular}[c]{@{}l@{}}Depression, PTSD,\\anxiety, ADHD\end{tabular} &
  \cellcolor[HTML]{FA8A72}11 &
  Family, professionals, social media, search engine \\ \hline
P16 &
  24 &
  Trans &
  College graduate &
  Black &
  Depression &
  \cellcolor[HTML]{FFEB84}8 &
  friends, professional, stranger \\ \hline
P17 &
  21 &
  M &
  High school &
  Black & no
   & 
  \cellcolor[HTML]{B1D47F}6 &
  \begin{tabular}[c]{@{}l@{}}Family, friends, professionals, social media, search engine, \\ teletherapy, services in the local community\end{tabular} \\ \hline
P18 &
  22 &
  M &
  College graduate &
  White &
  no &
  \cellcolor[HTML]{FCAA78}10 &
  Family, friends, social media, mobile apps, services in local communities \\ \hline

\end{tabular}%
}
\footnotesize Notes: PTSD: Post-traumatic stress disorder; ADHD: Attention-deficit/hyperactivity disorder
\end{table}

\subsection{Interview and Visual Elicitation}
The interviews averaged a duration of 1 hour and 20 minutes and each interviewee was compensated with a \$30 Amazon gift card. The interviews took place virtually via Zoom from January to September 2023. Before each interview, we introduced the scope of this study and answered any questions concerning the informed consent form. We also assured the participants that the interviews would be anonymous, confidential, and non-judgmental, encouraging them to be as open as they wanted.

We began the interviews by asking participants' current jobs, residential areas, and other background information. Then, the interviewer asked participants to describe their current mental health status. Example questions included, “Could you tell me about your mental health concerns?” “When did you first notice that?” and “How did it evolve over time?” 

\begin{figure}[h]

% \begin{subfigure}{0.3\textwidth}
% \centering
% \includegraphics[width=\linewidth]{images/Resources.png}
% \centering
% \caption{Visual prompts during interviews}
% \label{fig:visualprompts}
% \end{subfigure}
% \begin{subfigure}{0.6\textwidth}
\includegraphics[width=.8\linewidth]{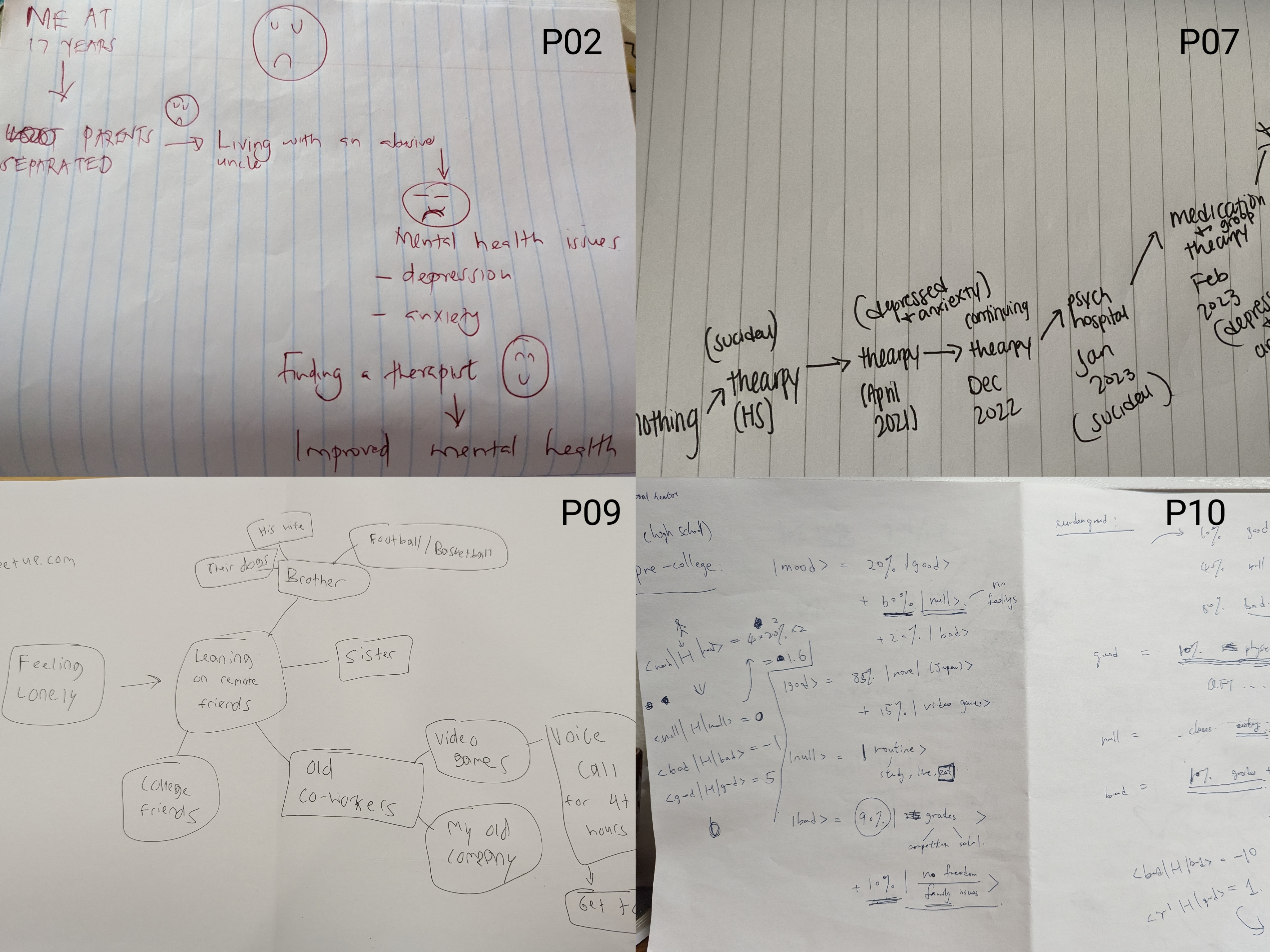}
\caption{Examples of Journey Maps Drawn by Participants. P02 and P07 presented their help-seeking timeline, P09 demonstrated his resources network, and P10 quantified his satisfaction with each resource using self-defined mathematic formulas.}
\footnotesize
\Description{
 Examples of Journey Maps Drawn by Participants. P02 and P07 presented their help-seeking timeline, P09 demonstrated his resources network, and P10 quantified his satisfaction with each resource using self-defined mathematic formulas.}
\label{fig:examplemaps}
% \end{subfigure}

% \caption{Interview materials}
% \label{fig:interviewmaterials}
\end{figure}

We utilized \textbf{visual elicitation techniques} for a richer exploration of their (non-)help-seeking journeys \cite{chen_timeline_2018}. When participants started to dive into specific story-telling, we asked them to draw their experiences of when, where, and how they sought help. Participants were encouraged to take some time to recall their experiences and present their stories in any visualization format. Our participants drew their help-seeking experiences from different perspectives, shown in Fig \ref{fig:examplemaps}. The interviewer then asked follow-up questions based on the drawings to elicit details, for example, “What did you do to cope with [a specific event] in the drawing?” “Did you talk with any people (e.g., family, friends) or use any technology (e.g., social media groups, mobile apps, hotlines)?” and "Why did you decide to do that?" In this process, the interviewer paid special attention to how they perceived and used different resources to seek help. The interviewer also encouraged participants to continually add contexts to the drawings as they narrated. Participants were instructed to focus the camera on the drawing to share with the interviewer during this process. 

After the interview, we asked participants to reflect on their experiences and evaluate their satisfaction with each mentioned resource using a five-point scale (1- very unsatisfied; 5- very satisfied). We then invited them to share the most satisfying and unsatisfying experiences, the challenges they encountered during help-seeking, and the ideal help they wished to get.
% reminded them to keep developing the drawing prompts like "they are great visuals to understand your story. Can you keep drawing as you tell me?" 
%After going through the help-seeking experiences on the journey map, we screen shared a checklist that aggregated resources listed in the U.S. Surgeon General’s Advisory \cite{office_of_the_surgeon_general_osg_protecting_2021} to remind participants of their interactions with other resources. 

\subsection{Social Ecological Theory Informed Coding Framework}
\label{sec: analysismethod}
All the interviews were conducted by the first author and were audio recorded and transcribed. The interviewer wrote debriefs immediately after each interview. We used iterative open coding and axial coding methods \cite{corbin_basics_2014} to systematically analyze the transcripts. The analysis was assisted by using NVivo, a qualitative content analysis software. This iterative process continued until the twelfth interview, at which point the codes reached a stable state, forming the foundation of our initial open coding schema. The open coding schema includes codes about resource type (e.g., technologies, family, professionals), support type (e.g., informational support, distraction), and challenges (e.g., high-cost, hard to talk). 

We followed the Social Ecological Theory \cite{stokols_translating_1996} 
%to conduct \enquote{ecological analyses characterize environmental settings as having multiple physical, social, and cultural dimensions that can influence a variety of health outcomes.} 
and categorized these resources into four levels: \textit{technological}, \textit{interpersonal}, \textit{community}, and \textit{societal} levels. The technological level emerged as a new category through our data analysis, while the other three levels were previously discussed in existing literature \cite{stokols_translating_1996,thompson_social_1990,seligman_depression_1975}.

Using this framework, we began axial coding to reorganize codes and generate themes and sub-themes related to participants' practices and challenges in seeking help. The two authors convened weekly meetings to review new interview debriefs, reflect on emerging themes in coding, compare codes with existing literature, and document these reflections in notes and memos. Through discussions, we purposely selected the next interviewees to further test and refine the framework. For instance, since societal resources were less frequently mentioned by earlier participants, we deliberately chose participants who utilized societal resources, focusing on how they defined, accessed, and perceived various societal services in comparison to resources at other levels. The initial open code "support type" evolved into the theme "support mechanisms" with three subthemes in axial coding. In addition, we systematically mapped resources used by participants harnessing the Socio-technical Ecosystem Framework, identifying two types of support systems, various pathways, and barriers in help-seeking. 

\subsection{Ethical Concerns}
This study was approved by the University Institutional Review Board (IRB). To protect the safety of our participants, in the informed consent form and at the beginning of interviews, we reminded the participants that they should feel free to take breaks during the interview and that they could exit the interview at any time. During the interview process, we kept sensitive to participants’ emotional changes when difficult experiences were disclosed and checked whether they would like to continue when negative emotions were observed \cite{draucker_developing_2009}. After the interview, we provided a list of mental health resources to the participants for future use, which was considered helpful and appreciated by many participants. 

\subsection{Positionality} 
Our research on mental health help-seeking behaviors and resource utilization is part of a broader effort to support young adults from diverse backgrounds through innovative technological solutions and social support. This work is informed by the lead author's lived experience with mental illness and involvement in peer support groups for young adults facing mental health challenges.

\subsection{Limitations}
Our study only included 18 participants who were young adults based in the U.S. who were willing to talk about their mental health concerns and help-seeking processes. They might also have less self-stigma. Thus, we acknowledge that their help-seeking behaviors could not represent all young adults, particularly those who did not seek help or were unwilling to talk about their experiences. Future research can examine the non-help-seeking behaviors of young adults.

\section{Findings}

To understand how marginalized young adults navigate their care-seeking journey (RQ1), we mapped participants' interactions with healthcare resources, revealing complex relationships between marginalized identity, mental health understanding, and care-seeking behaviors. In examining how resources facilitate or hinder care-seeking (RQ2), we identified pivotal "care encounters" that can catalyze aspirations and more proactive care-seeking behaviors. We also analyze the tensions in care encounters that fail to transform marginalized young adults' aspirations and behaviors.

\subsection{Participants' Mental Health Care Resource Use Practices}

Our participants represent a diverse range of backgrounds and varying self-identified marginality. Seven participants identified as first-generation college students, eight identified as racial or gender minorities, seven reported coming from low-income backgrounds, and three reported other forms of marginality, such as disabilities or recent immigration. Notably, five participants reported experiencing intersectional marginality.

\begin{figure}[h]
    \centering
    \includegraphics[width=\linewidth]{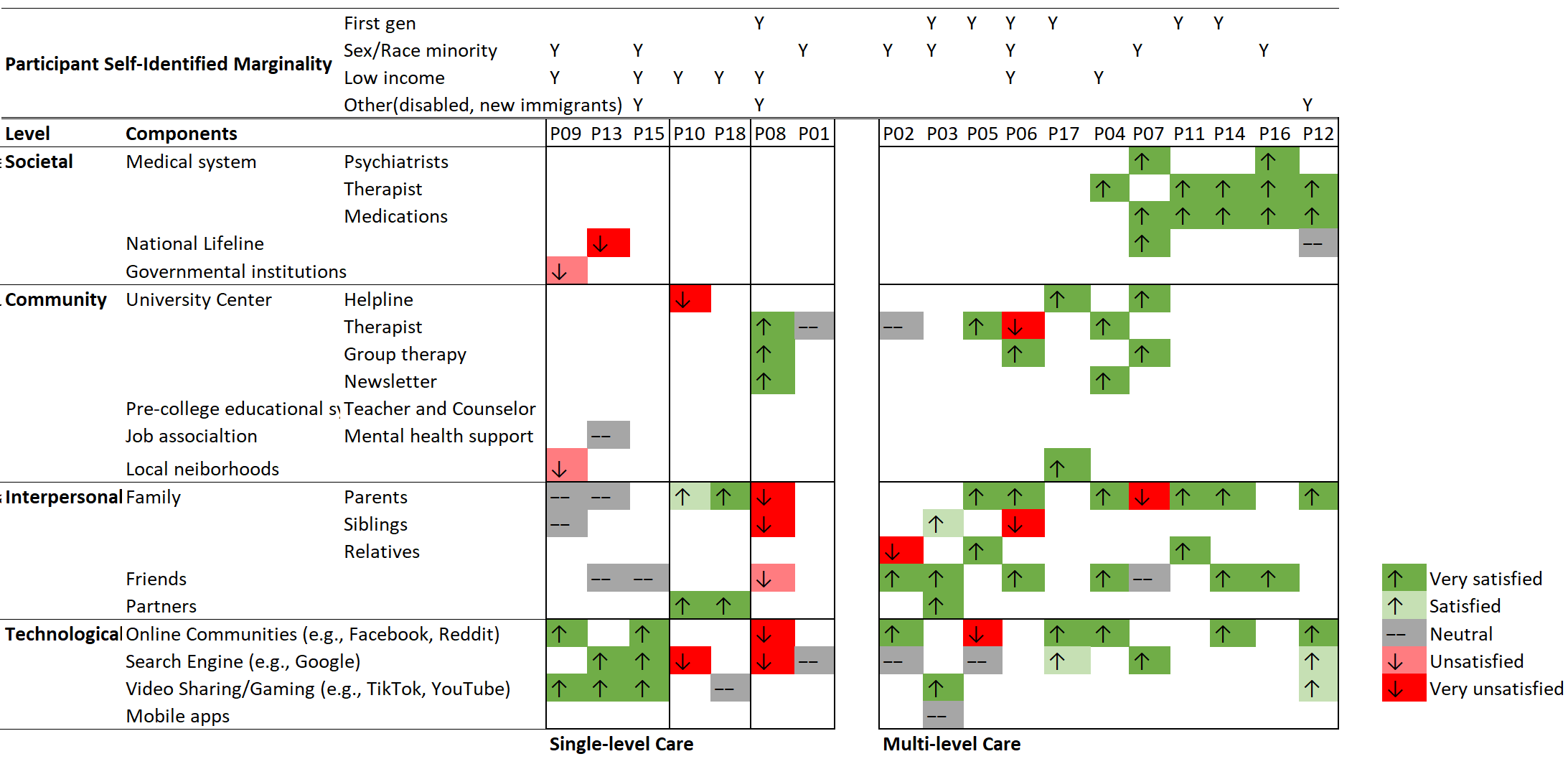}
    \caption{Participants' Interactions with Resources across Mental Health Care-Seeking Journey. Single Level Care denotes that participants have "satisfied" or "very satisfied" experiences with resources at one level; Multi-Level Care means that they have "satisfied" or "very satisfied" experiences with more than one type of resources.}
    \label{fig:supportsystem}
\end{figure}

All of our participants' care-seeking journey spanned more than a year and they shared both positive and negative experiences with different resources. In Figure \ref{fig:supportsystem}, we map their satisfaction with each resource they have interacted with based on their responses to the interview question: \enquote{To what extent are you satisfied with [a specific resource]?}

As mentioned in Section \ref{sec: analysismethod}, we followed the Social Ecological Theory \cite{stokols_translating_1996} and categorized these resources into four levels.
The \textit{technological level} includes search engines, online forums, video-sharing platforms, and mobile apps. These resources were frequently mentioned by participants and most of the use experiences were satisfying, but it should be noted that these resources had relatively small effects on participants. 
The \textit{interpersonal level} consists of participants' interpersonal ties. Except for P01, all other participants at least tried to seek help from family, friends, relatives, or partners, although six of them got neutral or unsatisfying feedback. 
The \textit{community level} contains all resources provided by entities that participants are members of, such as pre-college education institutions, universities, companies, and local neighborhoods. 11/18 participants tried community-level resources and six participants had varying degrees of satisfaction. 
The \textit{societal level} contains non-profit institutions and professional care including professional help from medical systems, the National Lifeline, and governmental institutions. Only 8 participants utilized these resources. 

%The aggregation of experiences presented in Figure \ref{fig:supportsystem} provides a static view and does not capture the temporal dynamics of participants' care-seeking journeys. 
As shown in Figure \ref{fig:supportsystem}, seven participants only received satisfying support from a single level of resources (i.e. single-level care). Among them, three  (P09, P13, and P15) primarily relied on technological solutions, two (P10 and P18) mainly sought emotional support from family members, and one (P08) only used resources from her community, the university mental health services. Notably, one participant (P01) did not report any significant experiences with care resources. The remaining eleven participants were able to navigate and obtain care from multiple levels of resources (i.e., multi-level care). We elaborate on participants' navigation of the socio-technical ecosystem in the following sections.

\subsection{Low Aspirations of Marginalized Young Adults' Care-Seeking}
\one{
Aspiration, understood as individuals' forward-looking capacity to aspire \cite{appadurai_capacity_2007}, shapes their engagement with care by influencing their behaviors and decisions \cite{pendse_mental_2019}. Our analysis reveals a pattern of low aspirations among participants, deeply rooted in their lived experiences of marginalization. These low aspirations manifest through internalized negative perceptions and strategic self-concealment, limiting their willingness to seek better care.}
%Our findings reveal how participants' aspirational capacity is diminished through recursive processes: lived experiences of marginalization (e.g., discrimination in healthcare settings, cultural dismissal of psychological distress) become internalized as diminished self-efficacy and expectations, 
\subsubsection{Lived Marginalization}

For many participants, marginalization extends beyond mental health concerns to encompass a broader complex interplay of cultural, personal, and societal factors where participants don't have access to various types of care resources. Some participants face extreme situations, as illustrated by P02:
\pquote{When I was 17 years old, my parents separated and I had no place to live. I was forced to live with my uncle, who was very abusive to me. It was at that point I developed some mental concerns because of the abuse... I have been detained, been spoken to [questioned] many times, and been denied food. It's just been so tough for me, you know? I just felt very alone. It was the lowest point of my life. It's just terrible. It's just awful. I was just like, The Dark Side of the Moon you know? (P02)} 
Others struggle with the unavailability of family support. P03 shared: \pquote{It's really busy with school and with family. My mom is struggling with back issues. My grandma, she has Psoriasis, so she has to get shots. My dad had heart surgery last summer. And then the only one that is currently able to help [me] is my eldest sister there, who is now a single mom starting to date a new guy and it's all a struggle. So I try, not to talk, to them. (P03)}

Social disparities they are experiencing further exacerbate these challenges, as expressed by P18, \pquote{I have a lot of friends that are rich, and they come from rich families. They go with themselves. But I can't join them because I always feel like I don't belong... discrimination and some other things might be part of the ceiling. 
So if one's information gets out, I think they'll use it again somewhere. (P18)} He was thus afraid of sharing his depression experiences with his "rich friends."

\subsubsection{Internalized Devaluation of Care}
\one{
Our analysis revealed a prevalent lack of aspiration for care among participants, manifested through devaluation of mental health importance and skepticism toward professional interventions. Cultural factors significantly perpetuate these internalizations, as one participant explained: \pquote{Mental health isn't really a thing in Asian communities. It's not really like shown by parents. I'm sure that there was one time when I was younger that I was sad and then my mom or dad was just like, 'deal with it by yourself.' And then that's what I do. I deal with it internally. (P01)}}

\one{
Beyond cultural influences, participants' low aspirations were further reinforced through their encounters with care institutions. These diminished expectations frequently stemmed from deep-seated disbelief in institutional care systems or negative organizational experiences. P02 expressed skepticism regarding governmental resources: \pquote{because you don't see these things are gonna happen. This is from the government, so the question is that these resources are very limited and you don't think you probably can get this. (P02)}
The limited accessibility of institutional information creates a self-reinforcing cycle. When institutions fail to provide accessible information, as P15 noted: \pquote{they [institutions] did not make the information accessible in a way that you can really get the support,} it confirms participants' existing skepticism and further diminishes their aspirations for care, perpetuating their disengagement from formal support systems.}

\subsubsection{Strategic Self-Concealment}
Our participants often conceal their mental health concerns and adopt a passive stance toward seeking help. They frequently "\textit{wait}" (twelve participants) for what they perceive as the "\textit{right opportunity}" (six participants) to address their mental health issues. This ambiguous pattern is common, as exemplified by P10, who expressed that he would treat his occasional depressive episodes more seriously "\textit{later}" but did not have a concrete plan, 
\pquote{I mean, I might tell my parents when it's the right time.}
Concealing mental health situations might stem from a view that mental health struggles are personal challenges to be overcome independently, as P03 articulated:
\pquote{I think it really also is important to have sort of a self-serving with mental health because you're with yourself for the rest of your life, right? Yes, it's important to have a support system externally, but internally is more important. (P03)}

%Although passive, such behavior also suggest they are still hopeful for the care that not yet come.
%Some have more positive \pquote{I haven't found a therapist that's a good fit, but I'm hopeful. (P15)}

\subsection{"Care Encounters" as Moments of Aspirational and Behavioral Changes}
\label{sec: encounterfindings}
While participants' care-seeking behaviors may appear passive, they reflect a nuanced approach of receptive anticipation rather than active avoidance of care resources. Our analysis revealed "care encounters" - serendipitous moments when participants discovered or were exposed to previously unimagined resources. We identified three distinct types of encounters in Figure \ref{fig:careencounter} that transformed participants' relationships with care: tangible assistance (e.g., free therapy sessions, academic accommodations, assisted hospitalization), supportive discourse (e.g., normalized conversations about mental health in both community contexts and intimate interpersonal exchanges), and social connection building that fosters meaningful relationships. These encounters helped marginalized young adults expand their aspirations for care and overcome their initial resource limitations. We further examined the affordances of the resource ecosystem that facilitated these transformative encounters.

\begin{figure}[h]
    \centering
    \includegraphics[width=0.6\linewidth]{images/careencounter.jpg}
    \caption{Mechanisms of Care Encounters}
    \label{fig:careencounter}
\end{figure}

%We found that in participants' interactions with the care resource ecosystem, there are opportunities that the ecosystem facilitate the aspirations, create open discourses about mental health, and foster supportive social groups that  We elaborate institutional agency and/or individual's agency interact in this process and identify the mechanisms.

\subsubsection{Encountering Tangible Assistance}
The most direct and impactful encounters that reshape participants' perceptions and behavior of care-seeking occur through tangible assistance, such as free therapy sessions, performance accommodations, and hospitalization services. We identified two circumstances that facilitate participants' engagement with tangible assistance: accessibility-driven encounters, where designated resource availability motivates engagement, and emergency-driven encounters, which occur during crisis situations.

\paragraph{Designated Accessibility}
Beyond mere availability, designated accessibility refers to how care resources thoughtfully woven into the fabric of participants' daily lives became more inviting and attainable. P08's experience exemplifies this transition in perception and behavior. During the isolating period of the COVID-19 pandemic, a series of email reminders about depression awareness became her unexpected lifeline:
\pquote{I think it was like halfway through Covid, I received consecutive emails day by day, saying to focus on my mental health. That was when I felt all the stressors were hitting on me all at once and I wondered why didn't I have a break from school? Let me, just, attend one of these Zoom Meetings, and see what they are about. Let me just check these resources out. It was during that time that I realized that I probably needed to get some help because these are the symptoms.}
Following these initial encounters, P08 utilized the free therapy sessions introduced in the workshops and emails—a pathway also taken by four other participants. 

For instance, P06 benefited from a reduced class load; P07 was offered extended tuition payment options; and P08 maintained access to free therapy sessions even after graduation. These accommodations strengthened the incentive for care-seeking, demonstrating that inclusiveness extends beyond vague promises. Reflecting on this experience, P08 expressed surprise at the abundance of available resources: \pquote{\elide{I think I'm more familiar with the [university] resources, and }I wasn't previously aware that there are so many resources out there possible to help you, and was just waiting for, like reach out and stuff, until I was told I was offered it.}

%For instance, P12's experience illustrates the critical importance of availability in moments of crisis. He turned to the national lifeline when no other ones seems possible, \pquote{I just wanted to talk to someone that would listen to me... I think I got what I needed out of it, like just talking to someone because it was able to calm me down quite a bit.} 

\paragraph{Responsive Interventions as Emergencies Arise}
Responsive interventions, which refer to situation-triggered responses to acute mental health needs, play a crucial role in challenging and changing participants' biases or negative attitudes toward care-seeking. These interventions are particularly vital during emergencies such as suicidal ideation or severe mood episodes.
P07's experience with the university crisis line exemplifies the comprehensive approach taken by community resources. Upon learning about her self-harming tendencies, the staff mobilized a multi-faceted response: \pquote{I stayed in the hospital for two days...they diagnosed me with a major depressive disorder and generalized anxiety disorder, and then that's when they gave me medications to start to treat both of those.} This intervention not only provided immediate care but also set P07 on a path of ongoing treatment and support. It demonstrates how active interventions can bridge the gap between the awareness of resources and the actual engagement with mental health services, especially in critical situations.

Besides the community-provided mental health services, P16's story illustrates the proactive role of interpersonal relationships during a crisis. After both of her parents passed away, P16 isolated herself at home for two months, unaware that she might be experiencing depression. Her friends' persistent check-ins culminated in a crucial intervention:
\pquote{This was too much for me. They came in when I lost it during that time. They took me to the hospital, checked with the hospital, and showed that I really needed this [professional help]. Then they decided that I needed to change the environment and of course, I listened to them. They are my support network and I tell everyone to find your network. This is what truly supports you no matter what bad happens.}

\subsubsection{Encountering Supportive Discourses}
Supportive discourses can open conversations and messaging that normalize mental health discussions and validate care-seeking behaviors. Participants frequently mentioned encountering these conversations in spaces that created supportive environments for mental health narratives, which, often gradually, mitigate their perceived stigmas of mental health and facilitate their further engagement with mental health resources.

\paragraph{Inclusive Atmosphere}
As mentioned in Section 4.1, many participants faced stigmas due to their familial, cultural, or religious backgrounds. However, some eventually encountered inclusive environments and spaces that actively welcomed conversations around mental health experiences.

The university campus was frequently mentioned as a microcosm of this awareness-raising ecosystem. For instance, P10 and P13's observations of omnipresent hotline signs in hallways, offices, and dining halls paint a picture of an environment saturated with care-seeking encouragement. Additionally, participants reported finding such spaces across various channels, from digital platforms like emails (P08, P16) and social media (P11) to physical spaces hosting awareness events (P15, P06). These messages create a constant, gentle nudge toward mental health consciousness, normalizing conversations around mental health and subtly reinforcing the availability of support.
These inclusive encounters also facilitated discussions that were previously deemed impossible. P07, unable to grieve her grandfather's death within her family setting, finally found an opportunity to process her grief in a university event that had been held off since middle school.
%The process of encountering and embracing these inclusive spaces often took time, with participants' attitudes changing after multiple exposures. 
%P07's journey with medication exemplifies this process. Repeated encounters with recommendations from online searches gradually shifted her perspective, leading her to consider medication as a potential solution for her mental health concerns. 

%The role of family and friends in identifying early-stage symptoms cannot be overstated. Their daily interactions provide a unique vantage point for recognizing subtle changes in behavior or mood. For five participants in our study, this intimate awareness led to early interventions. The experiences of P07 and P16, who were taken to psychiatrists by their parents during elementary school, highlight the critical role of familial support in initiating professional care.

\paragraph{Intimate Conversations}
One major reason behind the waiting and concealing behaviors among marginalized young adults, mentioned by six participants in our study, was the challenge of initiating open conversations about mental health with family and friends. We observed a delicate tension between their desire for support and the difficulty in articulating their needs. P13's reflection captures this struggle: \pquote{I mean just it's hard to talk about it...I'm not good at talking to people about the hard stuff.}
In such situations, participants found breakthrough moments when family members and friends took the initiative to start these deeper conversations. These proactive gestures often opened doors to meaningful support. P14 shared how her mother's willingness to discuss her depression not only aided her recovery journey but also strengthened their relationship: \pquote{She was like, I was really scared you would do something... I feel like if she ever saw me in such a sad place... that she can understand, she knew about my situation, at least the general situations.}
%However, participants sometimes strategically chose partial disclosure of their true state. P14's perspective encapsulates this nuanced approach:\pquote{Obviously they [my friends] are gonna support me,}Yet, in the same breath, P14 admits, \pquote{actually I don't want to tell [them about] my mental distress.} This approach allows individuals to modulate their level of openness, sharing only what they feel comfortable revealing in any given situation. It's a delicate balance between seeking support and maintaining personal boundaries.

\subsubsection{Encountering New Social Connections}
Another crucial type of encounter involves the fortuitous discovery of people who truly empower care-seeking. These relationships form a vital layer of care, providing comfort and understanding in navigating mental health challenges.

\paragraph{New Friends}
For many marginalized young adults, pivotal moments arise when they connect with peers, friends, or even strangers who offer transformative support, both immediate and long-lasting. P12 found solace in a therapist whose care evolved into a nurturing paternal connection: \pquote{I feel connected with him. He talks to me like a daughter because he has a daughter like me. I enjoyed going to play with his family and his children.} P15's journey was transformed by a sponsor he met at a peer-support group who became like family, extending both emotional and financial support for treatment. Sometimes, the most profound connections emerge from chance encounters, as P16 discovered during a moment of crisis outside a shopping mall: \pquote{She's a stranger, total stranger... I was just getting out of the shopping mall. That's when I broke down. She was like, oh, what's wrong? Can you sit down?
I cried a lot, cried, cried and I was talking and talking and talking and she was there she listening... I felt really very free with her because she was, she gave me, that, atmosphere for listening.}

\paragraph{Online Communities}
The development of supportive relationships extends beyond immediate circles to online communities, where individuals find connections based on shared experiences. The digital landscape further expands these information-sharing possibilities. Online communities provide platforms for reading others' stories, commenting on posts, and sharing personal experiences. For some, like P05 and P18, these "online friends" even surpass "real friends" in terms of support and understanding. P02's engagement with Facebook groups, where she formed meaningful connections with two online friends, exemplifies this phenomenon. These digital spaces become forums for direct exchange of information about symptoms, potential causes, and coping strategies.
The value of these online communities is particularly evident in P12's reflection:
\pquote{I will say this online group is safer because the online friends have seen their issues but, the majority of my physical friends don't experience what I'm experiencing.}
This sentiment underscores the unique comfort found in communities of shared experience, where understanding comes from firsthand knowledge of similar struggles.

\subsection{Aspirational Tensions in Care Encounters}
However, we also observe aspirational tensions in care encounters that fail to transform marginalized young adults' care-seeking motivations and behaviors. These tensions arise when structural barriers and relational mismatches conflict with participants' hopes for meaningful support. We first summarize the pragmatic challenges at each level in Table \ref{tab:challenges} and demonstrate how these barriers disrupt aspiration cultivation and empowerment development in this process.
\label{sec: challenges}


\begin{table*}[h]
\caption{Challenges of "Care Encounters" at Each Type of Resources}
\label{tab:challenges}
\footnotesize
\begin{tblr}{
  width = \textwidth,
  colspec = {|Q[r,0.6]|Q[l,0.7]|Q[l,3]|}, 
  hlines,
  row{1} = {font=\bfseries,c,gray9},
}
Resource Category & Challenges & Examples \\ 
\SetCell[r=3]{m}{Technological} &  Aversion of technology & \pquote{The search engines, these, online things are too general. They are helpful when you search for a road trip, but it's not the mental health thing} (P10).  \\ 
& Emotional frustration & \pquote{having to dig for it can be a little bit demoralizing. I just want instant gratification, but when your brain's all cluttered, the last thing you really want is to have to fight for help.} (P15)\\
& Lack experience-based personalization & \pquote{I get more worried when I look for it because there's just so much information out there... this information on Google might just be like for everybody. I get more anxious because I can't find information that is tailored to my needs.} (P08)\\
\SetCell[r=1]{m}{Interpersonal} & Difficulties in disclosure &  \pquote{My family don't really talk about things [mental health related] like that. } (P03) \\

\SetCell[r=3]{m}{Community} 
 & Varying professionality & “there are so many people talking while they are not qualified to talk” (P14) \\
  & Temporality & \par \pquote{It can be hard when you go from university to an external institution because you have to find one that matches your insurance. The first time, I called at least 10 or so people before I got the practitioner. They were like, 'We are not taking patients or we're not taking certain types of patients.' The whole process can be lengthy and challenging}. (P01) \\ 
  & Varying service investment & P15's university provides free and unlimited therapy sessions, while P07's university imposes a three-session limit. \\
  
\SetCell[r=2]{m}{Societal} & Compatibility & \pquote{the first one I had was just really weird and would like tell me her personal problems. And then the second one... she wasn't helpful.} (P06) \\
 & Impersonal & \pquote{She kept talking about sadness and stuff. And I tried to tell her, I'm not sad, I'm tired. I stopped seeing her because I felt like she wasn't really listening to me but just reacting.} (P01) \\
\end{tblr}
\end{table*}

\subsubsection{Identity Mismatch and Breakdown of Aspiration}
Our analysis shows that identity mismatches between marginalized individuals and mental health resources frequently erode aspiration before engagement begins.
Participants emphasized that aspiration for care-seeking depends not only on service availability but on whether providers reflect their cultural, racial, and lived experiences.
P01's search for a therapist exemplifies this dynamic: \pquote{I tried to find a therapist who is also a black woman, but I did not see any on the website....I don't think a male would understand [my experiences].} The scarcity of providers who shared her identity mattered beyond personal preference—it signaled that her specific needs and experiences might not be understood or respected, fundamentally undermining her motivation to pursue care. Similar sentiments emerged across participants with intersectional marginalized identities, who saw the lack of representation as diminishing their hope for meaningful therapeutic connection and effective support.

This erosion of aspiration through identity misalignment extended to peer support spaces. Online mental health communities, while intended to connect people with shared struggles, could deflate participants' motivation when dominant narratives failed to resonate. P05's account illustrates this gap: \pquote{I'm sorry, mine [situation] was going up for two or three weeks. Most of them are like coincidences. So I feel like I was tragic and traumatized because I was being attacked and my friend was beaten to death.} When confronted with posts about relatively minor stressors while processing severe trauma, participants like P05 reported feeling their aspirations for peer support collapse rather than flourish. Such experiential mismatches undermined the sense of possibility these spaces promised and dampened motivation to engage with online communities.

Even when participants located potential resources, structural and design barriers often reinforced these aspirational deficits. P15, seeking free therapeutic services due to financial constraints, described how navigating poorly designed systems crushed her initial motivation: \pquote{Every time I've looked it up, it seemed inaccessible. Not every website is very user-friendly. You have to go digging and digging to find what you need to find. And when I'm frustrated and feeling like I need help, that's the last thing I want to do.} She further noted that search results often prioritized profit-driven organizations over community resources: \pquote{Results from Google are not mainly from your local clinics. It's people that pay more money for their website to show up first... But they just look like businesses.} Such experiences fed the perception that care systems prioritized commercial interests over genuine understanding, systematically deflating aspirations for meaningful, community-centered support.

\subsubsection{Tangible Resource Gaps and Diminished Aspiration.}
Our analysis reveals how different types of mental health resources vary in their ability to provide tangible assistance that lifts and sustains aspiration among marginalized young adults in their care-seeking journey. 

Technological resources, despite their widespread use among participants, demonstrate the most limited capacity for tangible lift. While participants frequently initiated their help-seeking journey through technological platforms for information gathering and peer connection, these resources primarily offered informational rather than material support. Notably, four participants who initially relied exclusively on technological resources ultimately transitioned to other forms of support when they required concrete assistance beyond digital connection. This pattern indicates that while technology can facilitate initial aspiration through accessible information and community building, its inherent limitations in providing tangible interventions—such as direct therapy, crisis intervention, or material accommodations—constrain its capacity to sustain empowerment in mental health care-seeking.

Community-level resources present variable capacity for tangible support that directly impacts aspiration development. Some community entities actively provide concrete assistance through designated accessibility measures, while others offer limited material support despite good intentions. For instance, P13, a sports official, described the absence of tangible workplace mental health resources: \pquote{There's no training on how to handle the stress... we're just expected to deal with it.} Conversely, P10's institution provided sustained tangible support through frequent mental health awareness events and accessible counseling services, creating an environment where aspiration could flourish through concrete resource availability. This institutional variability in tangible support creates uneven landscapes for aspiration development, where individual empowerment becomes contingent on the material commitment of community institutions.

Societal-level resources, while possessing the greatest capacity for comprehensive tangible assistance—such as free therapy, emergency hospitalization, and systematic accommodations—often deliver this support through structures that diminish rather than cultivate aspiration. Only six participants successfully accessed these resources, with many encountering overwhelming procedural barriers that transformed potentially empowering tangible help into bureaucratic ordeals. As P08 explained: \pquote{Because I felt I could not handle it alone anymore and they were the closest and had anonymity.} This suggests that societal resources often become emergency-driven encounters accessed through crisis rather than accessibility-driven engagement, where the very systems designed to provide the most substantial tangible lift inadvertently undermine aspiration through their reactive, crisis-oriented delivery mechanisms.

%The temporal dimension of support emerged as a critical factor in sustained agency development. While community-level resources often provide proactive outreach, their temporary nature and position outside the professional medical system limits their ability to foster long-term agency in care-seeking. This temporal limitation can interrupt the development of sustained care-seeking behaviors, particularly for marginalized individuals who may require consistent support to build and maintain agency.

%The "official procedure" involved in seeking professional assistance also contributes to stigmatization, as articulated by P13, who described the difficulties , \pquote{And underneath the bridge over here is where all the mental health and all that building is. They have an office in here. But to get in there, you have to use this entrance, and this is the entrance, and these are the windows of our building.}

%\pquote{I just kinda, like, wanna keep it private. But at the same time, I know that people use it as a safe forum to discuss their issues and get comments and feedback. But I don't think it occurred to me at all to post my question. I don't feel comfortable releasing it to the public} (P08)

\section{Discussion}
\one{
This study reveals how marginalized identities intertwine with low aspiration in mental health care-seeking, extending beyond traditional resource barriers. We introduce the concept of "care encounters" to supplement the active assumptions underlying traditional "care-seeking" models. We conclude by discussing the implications of aspiration-centered design to carefully engineer "technology-mediated care encounters," arguing for shifting interventions toward low-tech solutions embedded within each individual's socio-technical ecosystem of care resources.}

%As shown in Figure \ref{fig:careencounter}, the 

%Characterizing Marginalized Young Adults Care-Seeking Behaviors

%\subsubsection{Integrating Identity Perception into Care-Seeking Frameworks}
\subsection{Marginalized Identity, Low Aspiration, and Mental Health Care-Seeking}
\one{
We contribute to mental health disparities literature by characterizing individuals with marginalized identities' care-seeking pattern of low aspiration, revealing how \textit{lived marginalization} and its \textit{internalized perception} operate recursively, producing strategic \textit{self-concealment practices}.}

\one{
Drawing on \textcite{appadurai_capacity_2007} conceptualization of aspiration as a forward-looking "capacity to aspire," we examine these navigational capacities as unequally distributed across social strata. Our findings highlight that marginality as an identity dimension shapes care-seeking through mechanisms exceeding resource limitations \cite{gulliver_perceived_2010}, functioning instead through internalized devaluation of care that loops into low aspiration. While structural inequities often constrain access to mental health care resources \cite{mani_poverty_2013}, we emphasize that low aspiration extends beyond mere external "barriers to care" as traditionally conceptualized in healthcare utilization models \cite{andersen_revisiting_1995}, such as financial constraints, insurance coverage, or time limitations \cite{mojtabai_treatment_2006, clement_what_2015}. This recurring process becomes internalized in identity formation.
Participants recalibrate expectations downward in response to perceived systemic exclusion \cite{yoshikawa_effects_2012}. Multiple participants demonstrated this by remaining inactive toward university and government mental health services despite their availability, with P12 dismissively noting, "\textit{it's something they [the university] have to do... I don't think they will be helpful}" — interpreting institutional support as merely performative.
This observation aligns with \textcite{pendse_marginalization_2023} proposition that resource scarcity reconfigures individuals' somatic understanding of mental health, suggesting marginality's influence shapes how individuals interpret symptoms, evaluate institutional legitimacy, and navigate care systems. }

\one{
This research advances existing models of mental health care-seeking behavior by positioning identity as a multifaceted dimension requiring greater attention in behavioral frameworks. Prior models \cite{davies_positioning_1990, major_social_2005, biddle_explaining_2007} and studies \cite{clement_what_2015} have identified "stigma" as a key barrier to care-seeking, capturing how the health-related identity of "being ill" hinders care-seeking \cite{biddle_explaining_2007}. However, our findings in Section 4.2 demonstrate that identity considerations operate through more complex mechanisms and reveal that identity is inherently multifaceted \cite{crocetti_multifaceted_2018} and intersectional \cite{schlesinger_intersectional_2017}, incorporating dimensions of social status and cultural background \cite{salgado_cultural_2015} that profoundly shape care-seeking decisions. For example, the socioeconomic dimension of identity observed in P17, who deliberately withholds depression symptoms from "rich" friends, suggesting how perceived socioeconomic differences create barriers to informal support networks.
These intersecting identities significantly impact individuals' aspirations regarding mental health care. For instance, P15, who identified as LGBTQ+, low-income, and disabled, described persistent feelings of "\textit{helplessness}" when navigating care resources when resources (e.g., government welfare) are theoretically accessible. Thus, we call for an expanded framework for understanding how identity processes shape mental health care engagement, particularly how intersecting marginalized identities may affect individuals' expectations and experiences when seeking mental health support.}

%Our study builds upon established literature by demonstrating how identity is intersectional \cite{schlesinger_intersectional_2017} and multifaceted \cite{crocetti_multifaceted_2018}, incorporating aspects of social status and cultural background \cite{salgado_cultural_2015} that significantly influence care-seeking decisions. Section 4.2 of our research identifies additional dimensions of identity that shape these processes. For instance, we observed socioeconomic dimensions of identity influencing informal support networks, as exemplified by P17, who deliberately withholds depression symptoms from "rich" friends.

%\subsubsection{Characterizing Low Aspiration and Dormant Hope} 
%Aspiration and critical theory. how it is defined. how does it overarchingly influence people in the margin?
%This behavior pattern is different from the barriers reported by prior lit such as indivisuals simply held back by financial or time barriers of seeking treatment. 
%evidence about internalized marginalization and low aspiration? 

However, our findings highlight opportunities for changing aspirations in mental health care engagement, as we observed that even when individuals demonstrate outward resistance, they frequently harbor "dormant hope" that aligns with psychological frameworks of readiness for change \cite{prochaska2005transtheoretical}. This differs distinctly from active resistance stemming from cultural or religious beliefs \cite{bhattacharjee_whats_2023}. Participant P06's experience illustrates this underlying receptivity: despite initially expressing reluctance, she scheduled therapy upon discovering accessible services. This pattern appeared among several participants, suggesting that strategic self-concealment often masks an openness to care that can be activated when nudged by external supports. The transformative power of such external nudges—including peers \cite{g_peer_2022, calear_sources_2022}, family \cite{musick_gaming_2021, theofanopoulou_exploring_2022}, and online information \cite{milton_seeking_2024}- has been reported in prior studies, particularly among marginalized populations \cite{anderson_cognitive_2005}. We argue that the dormant hope exhibited in ambivalent individuals opens up opportunities for interventions that restructure mental health resources and outreach efforts to better engage marginalized populations, which we will discuss in Section \ref{sec: designimplications}.
%Participants with unconcealed LGBTQ+ identity, experiences of homelessness, or unemployment often faced compound barriers due to disconnection from traditional support networks typically tied to stable affiliations,

\subsection{Rethinking Mental Health Care-Seeking Frameworks: Situating "Care Encounters"}
\subsubsection{"Care Encounters" as Catalysts of Aspiration Changes}

Our findings reveal a critical mechanism missing from current theoretical frameworks: serendipitous "care encounters" that serve as pivotal catalysts in reshaping marginalized young adults' care-seeking trajectories. These encounters differ fundamentally from traditional care-seeking in their motivational origin. Prior models utilized to explain mental health care-seeking behaviors, as identified in the literature review \cite{liu2025review}, typically begin with concepts such as "willingness" \cite{rickwood_when_2007}, "plan" \cite{ajzen1991theory}, and "autonomy" \cite{deci_support_1987-1}, assuming individual initiative drives care-seeking. Our findings, however, by centering on the experiences of marginalized young adults, challenge this assumption by demonstrating scenarios of low aspiration.

Most importantly, we demonstrate how these care encounters activate "dormant hope" and catalyze "aspiration changes" among marginalized young adults. Park \cite{pal_marginality_2013} conceptualizes marginality as a state of constrained agency and limited resources. Our findings show that these encounters create the conditions for motivation to emerge, potentially feeding into stronger aspirations and opening new opportunities for intervention. Our findings specifically illustrate how three mechanisms—tangible assistance, supportive discourses, and connection building—can transform care-seeking trajectories. Rather than presuming motivation as a prerequisite for engagement as traditional models suggest, this study provides implications for designing external circumstances that harbor meaningful care encounters.

This work echoes recent scholarship on \textbf{aspiration-centered design} \cite{freeman_aspirational_2017, pendse_mental_2019, kumar_aspirations-based_2019}, highlighting the importance of designing for aspiration formation, not just aspiration fulfillment. Kumar et al. \cite{kumar_aspirations-based_2019} elaborate that aspirations are influenced by sociocultural and environmental contexts and can change over time. Pendse et al. \cite{pendse_mental_2019} similarly highlight how mental health technologies must account for diverse aspirational contexts. Following this line of inquiry, our findings reveal how encounters can engineer aspirations in populations where mental health care-seeking may not initially feature in individuals' imagined futures. These findings contribute to understanding how to create transformative and empowering care ecosystems. Future studies can further develop the concept of encounter and its relationship with the infrastructure of individuals' care ecosystems, particularly examining how different types of encounters might dismantle challenges to aspiration formation among marginalized populations.
%While previous research has documented attitudinal changes toward care-seeking in digital spaces \cite{10.1145/3290605.3300294}, our study illuminates broader transformation patterns across both online and offline contexts. 

\subsubsection{Mapping the Socio-Technical Infrastructure of "Care Encounters"} 
To map the infrastructural enablers of care encounters within an ecological approach to care-seeking \cite{burgess_technology_2021, murnane_personal_2018, siddiqui_exploring_2023, ongwere_challenges_2022}, we employed Social Ecological Theory \cite{stokols_translating_1996} to examine participants' interactions with four distinct resource categories: societal, community, interpersonal, and technological as shown in Figure \ref{fig:supportsystem}. Our analysis reveals that the community sources emerge as principal contexts facilitating proactive care-seeking, while societal and technological resources often struggle to achieve similar engagement levels.

The ecological efficacy of community resources proves particularly powerful in empowering young adults with marginalized identities. Our findings highlight how university-based care outreach efforts foster empowerment through two interrelated channels: resource delivery providing accessible mental health support tools and information, while conversation invitations create safe spaces for open dialogue about mental health concerns. This dual approach proves particularly effective in cultivating supportive microenvironments that encourage care-seeking behaviors among marginalized young adults.

Community resources also serve as crucial bridges to societal resources while expanding marginalized young adults' interpersonal networks. Several participants, after positive experiences with university mental health services, decided to continue therapy post-graduation. For instance, P08's university extended her free therapy sessions while she searched for post-graduation care options. Communities can host events that foster interpersonal relationships, crucial for marginalized young adults who often experience exclusion from their families and lack friends comfortable with mental health discussions.
University-managed peer support groups emerged as critical spaces for connection. In one notable case, P17 encountered an alumnus who sponsored all treatment fees, including therapist visits and medication. Such local support groups were perceived by participants (e.g., P07 and P17) as more trustworthy and engaging compared to online peer support groups, enhancing the potential for positive care experiences \cite{gould_technology_2020}.
On the downside, it is crucial for future study to investigate support for individuals without readily available community resources. The four non-university-affiliated participants in our study—including individuals experiencing homelessness, unemployment, and workplace isolation—underscore the need for support systems beyond university contexts.

Our analysis reveals limitations in technological-level resources' efficacy in initiating active interventions and motivating care-seeking. P16's experience with online portals demonstrates how poorly designed technological interfaces can reinforce feelings of exclusion and diminish care aspirations. The limited engagement with online resources was also observed in previous studies \cite{pretorius_searching_2020}, suggests the disconnection of these tools from young adults' daily lives may explain their relatively low ecological validity \cite{mohr_three_2017}.
These findings emphasize the need for a more integrated approach to mental health support that combines technological solutions with community-based resources while recognizing the crucial role of interpersonal connections in facilitating care encounters.

%on online platforms or \cite{zhang_online_2018, metts_perceptions_2022}, and tracking apps \cite{costello_predictive_2020}.
%<Perceptions of Helpful and Unhelpful Responses to Disclosures of Suicidality in a Sample of Mobile App Users> 
%have highlighted the importance of resources aligned with the socio-economic situations and experiences of help providers to facilitate help-seeking \cite{gould_technology_2020}

% \cite{lattie_designing_2020} contextualized the needs of college students in their social ecosystem and social support networks and co-designed technologies that can be fitted into their everyday lives.  \textcite{le_exploring_2021} proposed a multifaceted approach incorporating mobile applications, individual interventions, and naturalistic conversations to mitigate college students' everyday anxiety. 
% Notably, a recent workshop emphasized an ecological perspective to enhance the accessibility of mental health services \cite{ongwere_challenges_2022}.
% <Conceptual framework for personal recovery in mental health: systematic review and narrative synthesis> The roles of pathways and resources \cite{leamy_conceptual_2011}.

\subsection{Engineering and Embedding Technology-Mediated "Care Encounters" into Young Adults' Socio-Technical Care Ecosystem}
\label{sec: designimplications}
\one{
Kumar et al. \cite{kumar_aspirations-based_2019} demonstrate that aspirations are fundamentally influenced by broader power dynamics and socio-structural factors that permeate communities, suggesting that embedding technologies is key for aspirations cultivation \cite{pendse_mental_2019}. Our findings align with this perspective—while many participants described their care encounters as "coincidence" or "luck," our analysis reveals these experiences are systematically enabled by infrastructural designs that structure and facilitate the possibility of such encounters \cite{pendse_like_2020}. We discuss the affordances of various resources in the socio-technical care ecosystem and identify critical design implications for improving mental health support systems.}

Building on research that integrates mental health technologies within young adults' social ecosystems \cite{lattie_designing_2020, le_exploring_2021, stefanidi_children_2023}, we propose several key implications to guide future researchers and designers in embedding and integrating technologies within marginalized young adults' social, cultural, and material environments \cite{gould_technology_2020}.

\subsubsection{Creating Care Encounters Through Low-Technology Outreach}

Despite extensive research on how people use mental health self-management applications \cite{wiljer_effects_2020, alqahtani_co-designing_2021}, our research revealed notably limited adoption of these tools among young adults with marginalized identities, with only two out of eighteen participants reporting attempted usage. This low adoption rate likely stems from insufficient trust and motivation \cite{burke_qualitative_2022}. However, our findings demonstrate that low-technology interventions can effectively create initial care encounters that serve as critical entry points into mental health support systems. Four participants specifically highlighted how university outreach through traditional channels—emails, posters, and social media—successfully initiated their first care encounters by making services visible and accessible \cite{gitlow_how_2019}. Similarly, three participants discovered the National Crisis Line through prominent search engine results, illustrating how strategic placement of information across commonly accessed platforms can generate spontaneous care encounters. These low-technology approaches function as effective catalysts for care encounters, requiring minimal technological literacy while maximizing reach and accessibility \cite{gould_technology_2020}. Rather than relying on sophisticated applications that may create barriers, these simple outreach mechanisms can serve as crucial first touchpoints that anchor critical information and foster gradual shifts toward help-seeking behaviors.

\subsubsection{Enhancing Care Encounter Efficacy Through Identity and Experience Alignment}
Our findings reveal that the quality and impact of care encounters improve when there is alignment between users' identities and lived experiences with those of care providers or peer supporters. Current platforms inadequately address this personalization need, particularly for marginalized users seeking meaningful connections with providers who share similar backgrounds. P05's difficulty relating to posts from more resourced individuals exemplifies how misaligned care encounters can diminish therapeutic value and reduce engagement \cite{liang_embracing_2021}. Participants demonstrated stronger engagement with care encounters when they encountered narratives and providers that mirrored their personal circumstances, cultural backgrounds, or shared experiences of marginalization. These findings suggest that effective care encounters are not merely about access to services, but about creating connections that resonate with users' authentic experiences. Future recommendation systems and online peer support platforms should prioritize identity-based and experience-based matching to enhance the efficacy of care encounters \cite{kim2020ever}. This approach to designing care encounters recognizes that therapeutic value emerges not just from professional expertise, but from the validation and understanding that comes through shared lived experiences, particularly among young adults navigating similar challenges \cite{de2016digital}.

\subsubsection{Facilitating Care Encounters for Deep Interpersonal and Family Conversations}
Most participants found it difficult to initiate conversations about mental health with family members and close friends. Specifically, five participants in our study experienced complicated family dynamics that created substantial obstacles to open dialogue and trust-building within their primary support networks \cite{chung_medical_2020}. These cultural and relational barriers highlight the critical need for alternative pathways to facilitate deep conversations that might otherwise remain impossible. Technology-mediated care encounters present unique opportunities to bridge these interpersonal gaps and create alternative support systems \cite{dalsgaard_mediated_2006, shin_designing_2021}. The emergence of empathetic chatbots and emotion-detection capabilities in Large Language Models \cite{wang_reprompt_2023} enables mediated care encounters that can facilitate difficult conversations and help young adults process complex family relationships \cite{jo_understanding_2023, koulouri_chatbots_2022}. Particularly noteworthy is how four male participants demonstrated the unexpected potential of gaming environments as venues for care encounters, with these digital spaces providing safe spaces for emotional expression and friendship maintenance away from family tensions \cite{wong-villacres_technology-mediated_2011}. These technology-mediated care encounters offer valuable alternative pathways for building intimacy when traditional face-to-face interactions prove challenging or emotionally fraught \cite{gould_technology_2020}.

%This aligns with recent research showing how peer support significantly enhances engagement with mobile mental health apps \cite{wong_postsecondary_2021, jonathan_smartphone-based_2021},

% \paragraph{Support Long-term Care-Seeking Journey}
% Eight participants' help-seeking journeys began through active intervention from interpersonal and community resources. This suggests an opportunity for technologies capable of monitoring early signs of mental health challenges to proactively engage independently living young adults. Future development should prioritize user involvement in early design stages \cite{alqahtani_co-designing_2021} and integrate innovative approaches such as asynchronous remote support \cite{bhattacharya_designing_2021} and multimodal communication \cite{h_barriers_2021} to enhance engagement with professional treatment.

\section{Conclusion}
Through interviews with 18 young adults with diverse marginalized identities, our study reveals how passive care-seeking behaviors are shaped by experiences of marginalization. We identify "care encounters" as critical turning points that can transform passive patterns into proactive engagement through tangible assistance, supportive discourse, and social connections. These findings inform the design of interventions that can intentionally facilitate such transformative encounters, ultimately improving marginalized young adults' access to mental health care.

%\begin{ack}
\section{Acknowledgement}
We extend our gratitude to the participants for their heartfelt sharing of their experiences and feelings, which has greatly inspired us. We also thank the reviewers for their valuable insights, which helped us refine findings and clarify contributions. 
This research is supported by the University of Texas at Austin School of Information John P. Commons Teaching Fellowship and the University of Texas at Austin Continuing Fellowship. We sincerely appreciate the feedback from Dr. Elliott Hauser, Dr. Andrew Dillon, and colleagues in the writing seminar Spring 2024 at UT Austin iSchool.
%\end{ack}

\bibliographystyle{ACM-Reference-Format}
\bibliography{references, extra}

\appendix
\end{document}